\documentclass[prl,twocolumn,superscriptaddress,showpacs,floatfix,longbibliography]{revtex4-1}
\usepackage{mathrsfs,braket}
\usepackage{amssymb, amsbsy, amsmath, latexsym, dsfont, array, layout,
graphicx,mathrsfs,color,ulem,bm}
\usepackage{cancel}
\usepackage[colorlinks=true,citecolor=blue,urlcolor=blue]{hyperref}

\newcommand{\one}{\mathds{1}}

\begin{document}

\title{Manipulating non-reciprocity in a two-dimensional magnetic quantum walk}

\author{Quan Lin}
  \affiliation{Beijing Computational Science Research Center, Beijing 100084, China}
\author{Wei Yi}\email{wyiz@ustc.edu.cn}
  \affiliation{CAS Key Laboratory of Quantum Information, University of Science and Technology of China, Hefei 230026, China}
 \affiliation{CAS Center For Excellence in Quantum Information and Quantum Physics, Hefei 230026, China}
\author{Peng Xue}\email{gnep.eux@gmail.com}
  \affiliation{Beijing Computational Science Research Center, Beijing 100084, China}

\begin{abstract}
\bf{Non-reciprocity is an important topic in fundamental physics and quantum-device design, as much effort has been devoted to its engineering and manipulation.
Here we experimentally demonstrate non-reciprocal transport in a two-dimensional quantum walk of photons, where the directional propagation is highly tunable through dissipation and synthetic magnetic flux. The non-reciprocal dynamics hereof is a manifestation of the non-Hermitian skin effect, with its direction continuously adjustable through the photon-loss parameters. By contrast, the synthetic flux originates from an engineered geometric phase, which competes with the non-Hermitian skin effect through magnetic confinement.
We further demonstrate how the non-reciprocity and synthetic flux impact the dynamics of the Floquet topological edge modes along an engineered boundary.
Our results exemplify an intriguing strategy for achieving tunable non-reciprocal transport, highlighting the interplay of non-Hermiticity and gauge fields in quantum systems of higher dimensions.
}
\end{abstract}

\maketitle

Open quantum systems are ubiquitous in nature, and exhibit rich and complex behaviors unknown to their closed counterparts~\cite{openbook}.
The recent progresses in non-Hermitian physics offer fresh insights into open systems from a unique perspective, giving rise to exotic symmetries and new paradigms of topology~\cite{benderreview,review2,photonpt1,review3,nonHtopo1,nonHtopo2,Lee,WZ1}.
A much studied non-Hermitian phenomenon of late is the non-Hermitian skin effect (NHSE)~\cite{WZ1,WZ2,murakami,ThomalePRB,stefano,tianshu,Budich,mcdonald,alvarez,lli,yzsgbz,stefano2,lyc,fangchenskin2,teskin,photonskin,metaskin,scienceskin,teskin2d,dzou,fangchenskin,kawabataskin}, whereby a macroscopic number of eigenstates become exponentially localized toward the boundaries. The NHSE has significant impact on the band topology~\cite{WZ1,WZ2,murakami}, the spectral symmetry~\cite{nonblochpt1,nonblochpt2,XDW+21}, and dynamics~\cite{quench1,quench2,coldatom}. One of the most salient dynamic signatures of the NHSE is the directional bulk flow~\cite{stefano,coldatom,ql,ql2}, which is closely connected to the global topology of the spectrum on the complex plane~\cite{fangchenskin}.
Such non-reciprocal dynamics can have potential applications in topological transport and quantum-device design, but the generation and control of this peculiar form of non-reciprocity, particularly in higher dimensions, remain experimentally unexplored.

In this work, we experimentally demonstrate the tuning of non-reciprocal transport in photonic quantum walks on a synthetic two-dimensional square lattice. The non-reciprocal dynamics underlies the NHSE of the two-dimensional quantum walk---the unidirectional flow leads to the accumulation of eigenstates toward boundaries in the corresponding direction. By tuning the photon-loss parameters, we show how the direction of the flow (hence the direction of the NHSE) can be continuously adjusted. The much discussed corner skin effect in two dimensions is but a special case here, corresponding to a directional flow along the diagonal of the square lattice.
By engineering the quantum-walk setup, we also introduce a synthetic flux to the lattice~\cite{hof,sai}, which we observe to suppress the non-reciprocal dynamics. Such a suppression is the result of the competition between two localization mechanisms: magnetic confinement and NHSE~\cite{chenwei}.
We quantitatively characterize the tunability of the non-reciprocity through loss and flux, and further demonstrate their impact on the dynamics of topological edge modes along the boundary.
Our experiment is the first observation of the magnetic suppression of NHSE, and further illustrates the flexible control over the NHSE-induced non-reciprocity in higher dimensions.

%given the extra spatial degrees of freedom,

%Just as the Bloch oscillation is a dynamic manifestation of the localized eigenstates of a Wannier-Stark ladder, the non-reciprocal transport is a dynamics signature of the NHSE.

\begin{figure*}[tbp]
\centering
\includegraphics[width=0.75\textwidth]{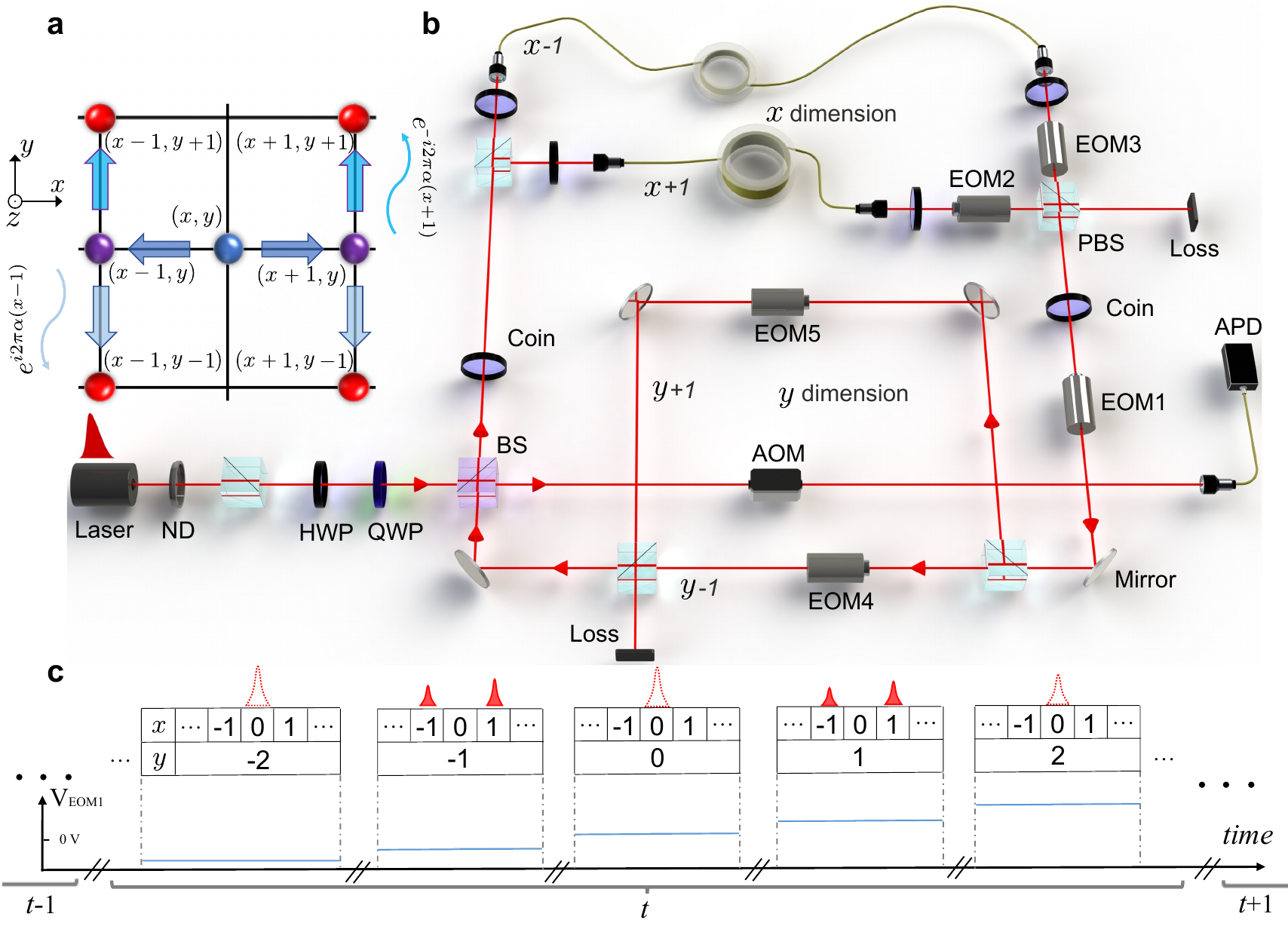}
\caption{{\bf Two-dimensional non-Hermitian quantum walk with a synthetic gauge field.} {\bf a} Schematics for possible movements of a walker at spatial position $(x,y)$ during each time step.
{\bf b} A time-multiplexed implementation of the two-dimensional photonic quantum walk. The photons are initialized at position $(0,0)$ in the superposition of the polarizations $(\ket{H}+i\ket{V})/\sqrt{2}$. Once coupled into the setup through a low-reflectivity beam splitter (BS, reflectivity $3\%$), their polarization state is manipulated by a half-wave plate (HWP). The photonic wave packets are split by a polarizing beam splitter (PBS) and routed through a pair of single-mode fibers (SMF) of length $287.03$m and $270$m, respectively, implementing a temporal step in the $x$ direction. A temporal step in the $y$ direction is implemented by another two-PBS loop based on the same principle, but in the free space instead of fibers. At each step, photons are partially coupled out to a polarization resolving detection of the arrival time via avalanche photodiodes (APDs). ND: neutral density filter; AOM: optical switch acousto-optic modulator; EOM: electro-optic modulator. {\bf c} Illustration of the operation sequence of the time-multiplexed quantum walk. Here $V_\text{EOM}$ is the control voltage applied to the EOMs.
}
\label{fig:fig1}
\end{figure*}

\begin{figure*}[tbp]
\centering
\includegraphics[width=0.67\textwidth]{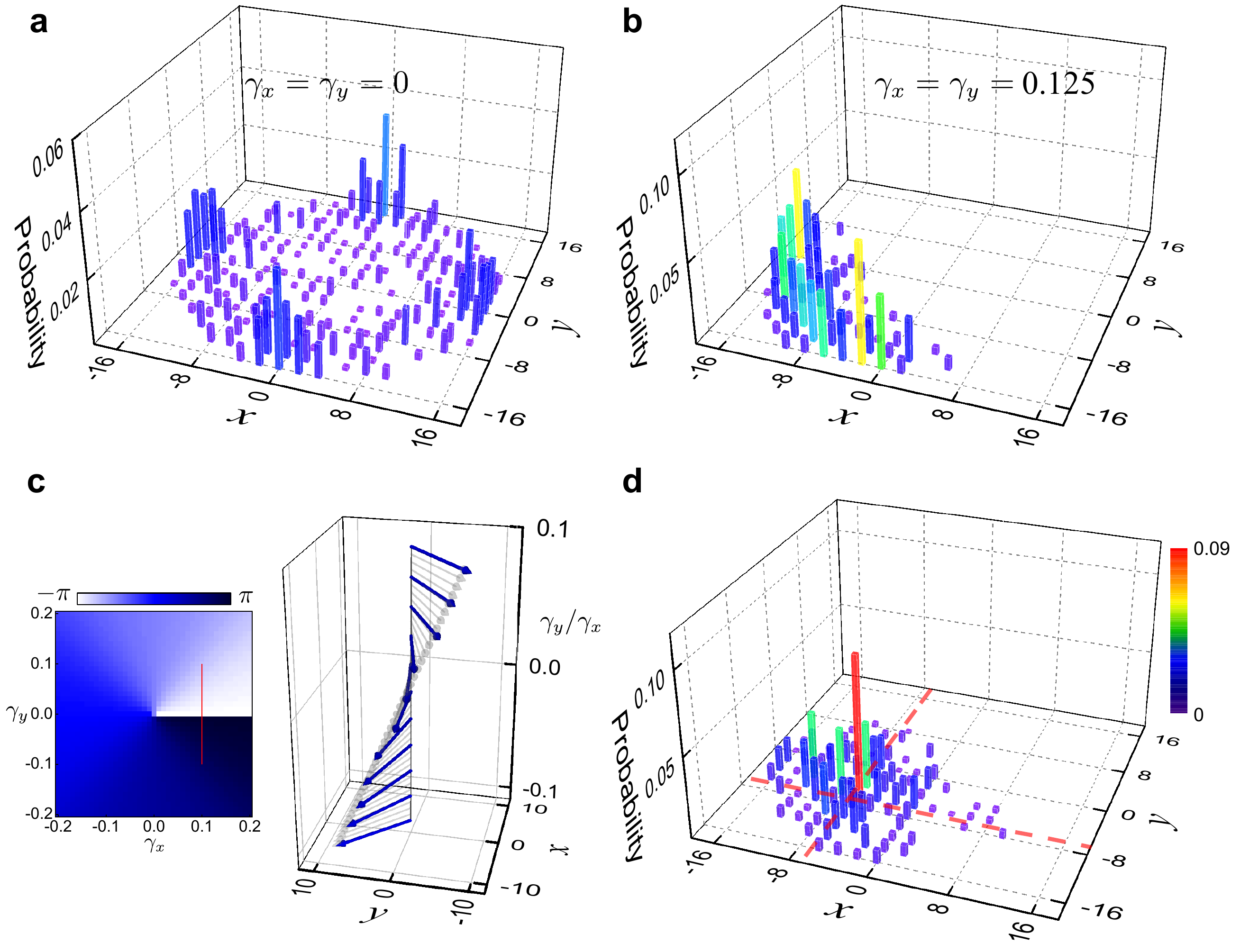}
\caption{{\bf Tunable non-reciprocity and NHSE.}
The walker with the polarization $(\ket{H}+i\ket{V})/\sqrt{2}$ starts from the lattice site $(0,0)$ with $\alpha=0$. Probability distributions are measured after $16$ time steps.
{\bf a} Probability distribution for a Hermitian two-dimensional quantum walk with $\gamma_x=\gamma_y=0$. {\bf b} Probability distribution following a non-Hermitian quantum walk with $\gamma_x=\gamma_y=0.125$.
{\bf c} Directional displacements after the final time step ($t=16$) for quantum walks with varying $\gamma_x $ and $\gamma_y$. (Left) Color contour of the azimuthal angle of the displacement $\bm{d}$ on the $x$--$y$ plane. (Right) Measured (blue arrows) and simulated (gray arrows) of the displacement along the red vertical line of the color contour (left panel).
{\bf d} Probability distribution following a non-Hermitian quantum walk in the presence of domain walls (marked by red dashed lines), with $\gamma_{x}=0.125$ for $x \geq-6$, $\gamma_{x}=-0.125$ for $x<-6$, $\gamma_{y}=0.125$ for $y\geq-6$, and $\gamma_{y}=-0.125$ for $y<-6$.
}
\label{fig:fig2}
\end{figure*}

\begin{figure*}[tbp]
\centering
\includegraphics[width=1\textwidth]{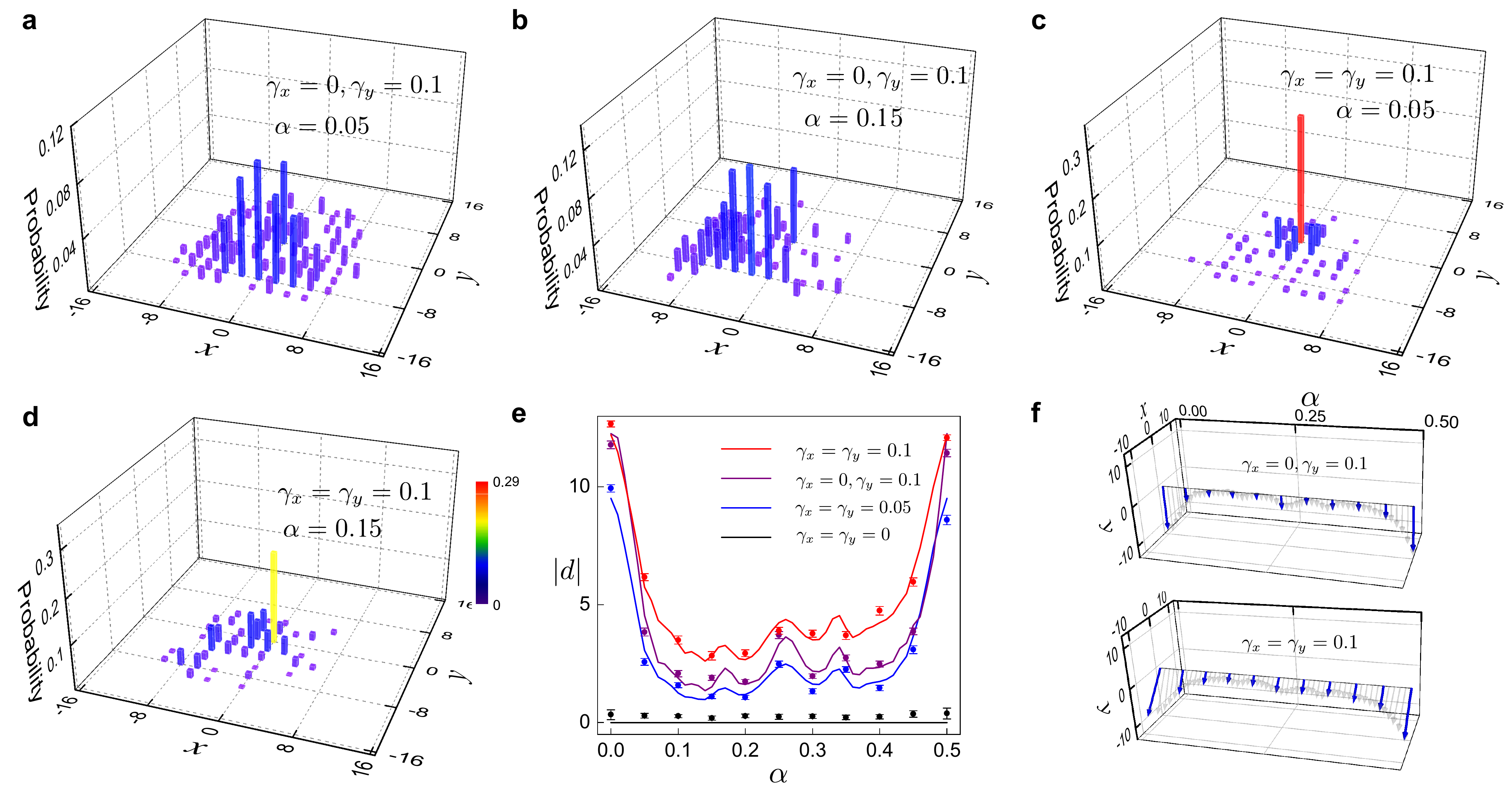}
\caption{{\bf Magnetic suppression of the non-reciprocal transport.} The walker is initialized in the state $\frac{1}{\sqrt{2}}(\ket{H}+i\ket{V})\otimes\ket{x=0}\ket{y=0}$. Measured probability distributions of $16$-time-step quantum walks with the tuning parameter $\alpha=0.05$, and the loss parameter $\gamma_{x}=0$, $\gamma_{y}=0.1$ in {\bf a}, with $\alpha=0.15$, $\gamma_{x}=0$, and $\gamma_{y}=0.1$ in {\bf b}, with $\alpha=0.05$ and $\gamma_{x}=\gamma_{y}=0.1$ in {\bf c}, and with $\alpha=0.15$ and $\gamma_{x}=\gamma_{y}=0.1$ in {\bf d}.
{\bf e} Norm of the directional displacement $\bm{d}$ for $16$-time-step quantum walks under different flux $\alpha$. Symbols represent the experimental data, and curves are the corresponding numerical simulations. Error bars are due to the statistical uncertainty in photon-number-counting. {\bf f} Measured (blue) and simulated (gray) $\bm{d}$. Blue arrows represent the experimental results, and gray ones indicate the corresponding numerical simulations.
%The similarities of the four cases are $0.929\pm0.005$, $0.938\pm0.005$, $0.922\pm0.006$ and $0.923\pm0.008$, respectively.
}
\label{fig:fig3}
\end{figure*}

{\bf Results}

{\bf Time-multiplexed two-dimensional quantum walk.}
In discrete-time quantum walks, the walker state $|\psi(t)\rangle$ evolves according to $|\psi(t)\rangle=U^{t}|\psi(0)\rangle$, where $t$ indicates the discrete time steps, and $U$ is thus identified as the Floquet operator that periodically drives the system. We consider such a quantum walk on a two-dimensional square lattice, with the Floquet operator
\begin{equation}
U=M_yS_yPCM_xS_xC.
\end{equation}
Here the shift operators are defined as $S_{j}=\sum_{\bm{r}}\ket{0}\bra{0}\otimes\ket{\bm{r}-\bm{e}_{j}}\bra{\bm{r}}+\ket{1}\bra{1}\otimes\ket{\bm{r}+\bm{e}_{j}}\bra{\bm{r}}$,
%$S_{y}=(\ket{0}\bra{0}\otimes\ket{y-1}\bra{y}+\ket{1}\bra{1}\otimes\ket{y+1}\bra{y})\otimes\one_{x}$, where $|x,y\rangle$ denote the position states,
with $\bm{r}=(x,y)\in \mathbb{Z}^2$ labeling the coordinates of the lattice sites, $j\in\{x,y\}$, and $\bm{e}_x=(1,0)$ and $\bm{e}_y=(0,1)$.
The shift operators move the walker in the corresponding directions, depending on the walker's internal degrees of freedom in the basis of $\{|0\rangle,|1\rangle\}$ (dubbed the coin states).
These coin states are subject to rotations under the coin operator
$C=\frac{1}{\sqrt{2}}\begin{pmatrix}1&1\\1&-1\end{pmatrix}\otimes\one_{\bm{r}}$,
where $\one_{\bm{r}}=\sum_{\bm{r}}\ket{\bm{r}}\bra{\bm{r}}$.
The gain-loss operators are given by (here $j\in\{x,y\}$)
\begin{align}
M_j(\gamma_j)= \begin{pmatrix}
                 e^{\gamma_j} & 0 \\
                 0 & e^{-\gamma_j}
               \end{pmatrix}\otimes\one_{\bm{r}},
\end{align}
which make the quantum walk non-unitary for finite $\gamma_x$ or $\gamma_y$.
%{\color{red}Note that $\gamma_x$ and $\gamma_y$ can in general be site dependent, though we fix them as constants unless otherwise specified. (deleted)}

A key ingredient to our scheme is the phase-shift operator, defined as
\begin{equation}
P=\sum_{\bm{r}}\begin{pmatrix}e^{i2\pi\alpha x}&0\\0&e^{-i2\pi\alpha x}\end{pmatrix}\otimes\ket{\bm{r}}\bra{\bm{r}},
\end{equation}
which enforces a position-dependent geometric phase on the walker, so that
the latter acquires a phase $2\pi\alpha$ when going around any single plaquette of the square lattice (see Fig.~\ref{fig:fig1}{\bf a}).
Similar to that of the Hofstadter model~\cite{hof}, the accumulated phase shift of the walker on the lattice is equal to the Aharonov-Bohm phase of a charged particle in a uniform magnetic field, with a magnetic flux $\alpha$ threaded through each plaquette. We therefore regard $\alpha$ as the synthetic flux, which takes value in the range $\left[0,1\right)$.

%For the particular choice of $U$ here, even and odd lattice sites are decoupled, leading to an effectively enlarged unit cell. It is thus sufficient to take the phase parameter $\alpha$ in the range $[0,0.5)$.

We experimentally implement the two-dimensional quantum walk above using photons. As illustrated in Fig.~\ref{fig:fig1}, the overall architecture is that of a fiber network~\cite{sch,ql,ql2}, through which attenuated single-photon pulses are sent, with each full cycle around the network representing a discrete time step. The coin states $\{|0\rangle,|1\rangle\}$ are encoded in the photon polarizations $\{|H\rangle,|V\rangle\}$.
The spatial degrees of freedom of the square lattice are encoded in the time domain, following a time-multiplexed scheme. This is achieved by building path-dependent time delays into the four different paths (labeled $x\pm 1$ and $y\pm 1$ in Fig.~\ref{fig:fig1}{\bf a}) within the network (see Methods for details).
The superpositions of multiple well-resolved pulses within the same discrete time step thus represent those of multiple spatial positions at the given time step (see Fig.~\ref{fig:fig1}{\bf b}).

The shift and coin operators are implemented with beam splitters (BSs) and wave plates (WPs), and the phase operator with one of the electro-optical modulators (EOM1 in Fig.~\ref{fig:fig1}).
We further implement polarization-dependent loss operators $M'_j=e^{-\gamma_j}M_j$ in each path, using a combination of the WPs and the EOMs. The time-evolved state driven by $U$ is then related to that in the experiment by adding a factor $e^{(\gamma_x+\gamma_y) t}$ to the latter.

For all experiments, avalanche photo-diodes (APDs) with temporal and polarization resolutions are employed to record the probability distribution of the walker states. This enables us to construct the site-resolved population of the synthetic lattice, with
\begin{align}
P_\text{exp}(x,y,t)=\frac{N(x,y,t)}{\sum_{x,y} N(x,y,t)},
\end{align}
where $N(x,y,t)$ is the total photon number on site $(x,y)$ at time $t$.

%\begin{figure*}[tbp]
%\centering
%\includegraphics[width=0.5\textwidth]{fig4}
%\caption{Effects on non-Hermitian skin effect in a two-dimensional quantum walk from the magnetic field. The walker with the polarizations $(\ket{H}+i\ket{V})/\sqrt{2}$ starts from the lattice site $(0,0)$. The boundary is set where photons can not reach. (a) Measured and simulated norm of displacement $\vec{d}$ for a $16$-step quantum walk with various $\gamma$. Symbols represent the experimental data, and curves are the corresponding numerical simulations. Error bars are due to the statistical uncertainty in photon-number-counting. (b) Measured and simulated $\vec{d}$ in spatial coordinates. Solid arrows represent the experimental results, and hollow ones indicate the corresponding numerical simulations.
%}
%\label{fig:fig4}
%\end{figure*}

\begin{figure*}[tbp]
\centering
\includegraphics[width=1\textwidth]{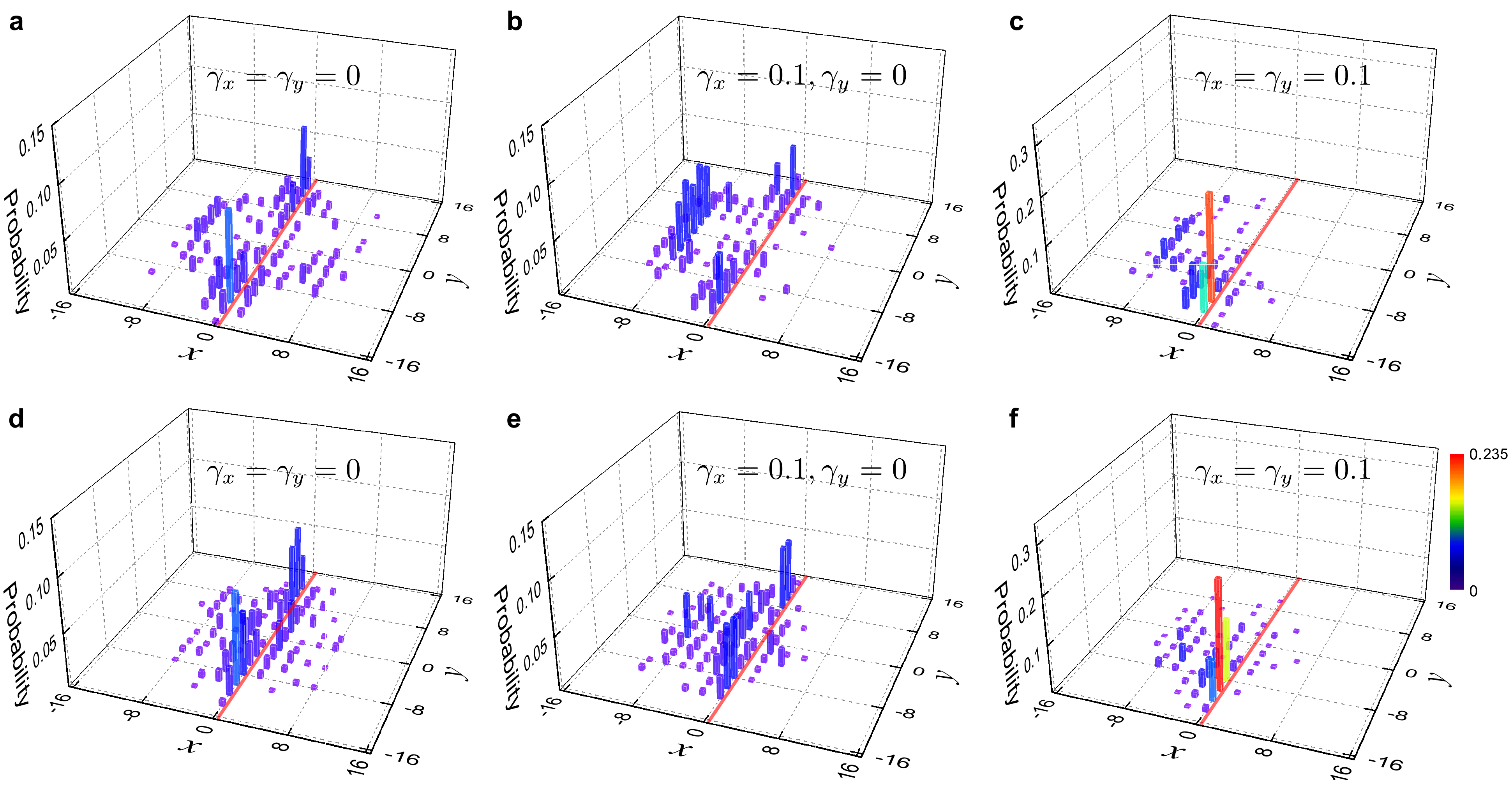}
\caption{{\bf Dynamic detection of topological edge modes in $16$-time-step two-dimensional magnetic quantum walks.} The walker is initialized in the state $\frac{1}{\sqrt{2}}(\ket{H}+i\ket{V})\otimes\ket{x=0}\ket{y=0}$ under a
domain-wall geometry. Specifically, in {\bf a, b, c}, we set $\alpha=0.05$ for $x\leq0$, and $\alpha=-0.05$ for $x>0$. In {\bf d, e, f}, we set $\alpha=1/3$ for $x\leq0$, and $\alpha=-1/3$ for $x>0$. The left, middle, and right columns show the measured probability distribution with the parameters $\gamma_{x}=\gamma_{y}=0$, $\gamma_{x}=0.1,\gamma_{y}=0$, and $\gamma_{x}=\gamma_{y}=0.1$, respectively.
%The similarities of the six cases are $0.930\pm0.004$, $0.941\pm0.004$, $0.946\pm0.004$, $0.955\pm0.003$, $0.951\pm0.003$ and $0.964\pm0.004$, respectively.
}
\label{fig:fig4}
\end{figure*}

{\bf NHSE and tunable non-reciprocity.}
In the absence of flux, quantum walks driven by $U$ already show non-reciprocal transport under
finite photon losses.
In Fig.~\ref{fig:fig2}{\bf a}, we show the measured populations of the synthetic lattice sites after $t=16$ time steps.
Starting from a local initial state at $\bm{r}=(0,0)$, the propagation in the synthetic spatial dimensions is symmetric along the four lattice directions (Fig.~\ref{fig:fig2}{\bf a}). However, under finite photon-loss parameters, the final-time photon distribution becomes asymmetric with a preferred direction. Such a directional flow is the signature of the non-reciprocal transport.
For instance, when $\gamma_x=\gamma_y\neq 0$, as shown in Fig.~\ref{fig:fig2}{\bf b}, the flow is diagonal to the square lattice.
By tuning the ratio of $\gamma_y/\gamma_x$, we can continuously adjust the direction of the asymmetric pattern. This is explicitly shown in Fig.~\ref{fig:fig2}{\bf c}, where we define the
directional displacement
\begin{align}
\bm{d}(t)=\sum_{x,y} \bm{r} P_\text{exp}(x,y,t).
\end{align}
As $\gamma_y/\gamma_x$ varies, the direction of the displacement at the final time step can be continuously tuned (see the left panel of Fig.~\ref{fig:fig2}{\bf c}). In our experiment, we adjust $\gamma_y/\gamma_x$ in the range of $[-1,1]$ for $16$-time-step quantum walks.
Correspondingly, the measured polar angle of $\bm{d}$ changes from $3\pi/4$ to $-3\pi/4$ (the right panel of Fig.~\ref{fig:fig2}{\bf c}).

%{\color{red}Because of the symmetry of our setup, the displacement can be tuned to any other directions on the $x$--$y$ plane, for instance, by reversing the flux.(deleted)}

Underlying this loss-induced directional flow is the NHSE in two dimensions. While it is straightforward to show that the direction of the non-reciprocal transport also indicates the direction of the eigenstate accumulation under the open boundary condition~\cite{supp},
from an experimental perspective, we observe the dynamic localization of the walker toward the boundary when a domain-wall boundary condition is imposed (see Fig.~\ref{fig:fig2}{\bf d}). Combined with the theoretical spectral analysis that there are no topological edge states present under the parameters of Fig.~\ref{fig:fig2}{\bf d}, it is clear that the localization is due to the NHSE.

{\bf Magnetic suppression of the NHSE.}
In Figs.~\ref{fig:fig3}{\bf a-d}, we show the final population distribution with the synthetic flux switched on, following $16$-time-step quantum walks. Compared to Fig.~\ref{fig:fig2}, the directional flow appears to be increasingly suppressed under larger $\alpha$, regardless of its direction. In Figs.~\ref{fig:fig3}{\bf e} and {\bf f}, we show the absolute values of the directional displacement $\bm{d}$ as functions of $\alpha$, for various loss parameters. The suppression is the largest when $\alpha$ is tuned in between $0$ and $0.5$. Such a suppression reflects the competition between the magnetic confinement and the NHSE, and can be used for the manipulation of the non-reciprocal transport.

{\bf Impact on topological edge states.}
In the absence of loss, the Floquet operator $U$ describes an anomalous Floquet Chern insulator, characterized by the Floquet topological invariant~\cite{sai,asb,rud}, which can be calculated for each quasienergy gap and is fully responsible for the topological edge states. Here we experimentally investigate how the NHSE under loss and magnetic confinement impact the topological edge states. For this purpose, we engineer a domain-wall configuration by choosing different values of $\alpha$ on either side of $x=0$.

As shown in Fig.~\ref{fig:fig4}{\bf a}, for lossless quantum walks, a pair of topological edge modes emerge, moving in opposite directions along the boundary. This is consistent with the prediction of the Floquet topological invariant (see Methods and \cite{supp}). When only the loss parameter $\gamma_x$ is turned on, the NHSE induces a horizontal directional flow toward the region with $x<0$. From the measured population following a $16$-time-step quantum walk (see Fig.~\ref{fig:fig4}{\bf b}), both the bulk flow and the topological edge states are clearly visible. Since the directional flow is perpendicular to the boundary, it has no direct impact on the motion of the topological edge states. This is no longer the case when both $\gamma_x$ and $\gamma_y$ become finite, as in Fig.~\ref{fig:fig4}{\bf c}. Here, besides a diagonal bulk flow induced by the corner skin effect, the topological edge modes moving in the negative (positive) $y$ direction are enhanced (suppressed) by the NHSE.

In Figs.~\ref{fig:fig4}{\bf d-\bf f}, we show the final probability distribution under a larger synthetic flux $\alpha$. Compared to Figs.~\ref{fig:fig4}{\bf a-\bf c}, the bulk propagation is significantly suppressed, whereas the topological edge modes are largely unaffected by flux. This suggests that magnetic confinement is helpful for the dynamics detection of topological edge modes in systems with the NHSE.

{\bf Discussion.}
We have experimentally demonstrated how the interplay of synthetic flux and dissipation enables the full control over the non-reciprocal transport underlying the NHSE. Since the quantum walk simulates an anomalous Floquet Chern insulator, we further illustrate how the motion of topological edge modes on the boundary is affected by the tuning parameters.
While the high tunability can be exploited for topological device design, our implementation of a dissipative anomalous Floquet Chern insulator further raises theoretical questions as to how the NHSE affects the bulk-boundary correspondence herein. Our experiment also paves the way for engineering more exotic forms of the non-Hermitian skin effect in higher dimensions~\cite{fangchenskin2} using quantum-walk dynamics.

%\clearpage

%\clearpage

{\bf Methods}

{\bf Experimental setup.}
We adopt a time-multiplexed scheme for the experimental realization of photonic quantum walks~\cite{sch,ql,ql2}.
As illustrated in Fig.~\ref{fig:fig1}, the photon source is provided by a pulsed laser with a central wavelength of $808$nm, a pulse width of $88$ps, and a repetition rate of $15.625$kHz. The pulses are attenuated by a neutral density filter, such that an average photon number per pulse is less than $2.4\times 10^{-4}$, which ensures a negligible probability of multi-photon events.
The photons are coupled in and out of a time-multiplexed setup through a BS with a reflectivity of $3\%$, corresponding to a low coupling rate of photons into the network. Such a low-reflectivity BS also enables the out-coupling of photons for measurement. A HWP with the setting angle $\pi/8$ is used to implement the coin operator $C$.

Four different paths in a fiber network correspond to the four different directions a walker can take in one step on a two-dimensional lattice. Two-PBS loops are used to realize polarization-dependent optical delays. The shift operator $S_x$ is implemented by separating photons corresponding to their two polarization components and routing them through the fibre loops, respectively. Polarization-dependent time delay is then introduced. Since the lengths of the two fiber loops are $287.03$m and $270.00$m, respectively, the time difference of photons traveling through two fiber loops is $80$ns.  The shift operator $S_y$ is implemented by another two-PBS loop based on the same principle, where the vertical component of photons is delayed relative to the horizontal component by a $1.61$m free space path difference. The corresponding time difference in the $y$ direction is then $4.83$ns.

The position-dependent phase operator $P$ is implemented using the first electro-optical modulator (EOM1).
The rise/fall times of EOM ($4$ns) are much shorter than the time difference between adjacent positions ($80$ns and $4.83$ns for $x$ and $y$ directions, respectively), which enable us to control the parameter $\phi$ precisely.

To realize a polarization-dependent loss operation $M'_x(\gamma_x)=e^{-\gamma_x}M_x(\gamma_x)$, two HWPs and an EOM are introduced into each fiber loop. Here HWPs are used to keep the polarizations of photons unchanged before and after they pass through the fiber loops. For $\gamma_x>0$, for the short loop, the voltage of EOM3 is tuned to $0$. Thus, after passing through the first PBS, horizontally polarized photons are all transmitted by the second PBS and are subject to further time evolution. Whereas for the long loop, by controlling the voltage of EOM2 to satisfy $\cos\theta/2=e^{-2\gamma_x}$, we flip part of the photons $(1-e^{4\gamma_x})$ with vertical polarization into horizontal ones. They are subsequently transmitted by the second PBS, and leak out of the setup. Otherwise, for $\gamma_x<0$, horizontally polarized photons are all transmitted by the second PBS and are subject to further time evolution in the long loop. By contrast, for the short loop, part of the photons $(1-e^{4\gamma_x})$ with vertical polarization are flipped by EOM3, transmitted by the second PBS and subsequently leak out of the setup. We use the same method to realize $M'_y(\gamma_y)$.

%The probability distribution of the walker $P_\text{th}(x,y,t)=\frac{|\langle x|\bra{y}\psi'(t)\rangle|^2}{\sum_{x,y} |\langle x|\bra{y}\psi'(t)\rangle|^2}$ at the position $(x,y)$ and step $t$ is obtained by dividing the number of photons collected by APD at the position $(x,y)$ using the total photons collected after the $t$th step, i.e.,
%\begin{equation}
%P_\text{exp}(x,y,t)=\frac{N(x,y,t)}{\sum_{x,y} N(x,y,t)}
%\end{equation}
%with $N(x,y,t)=N_H(x,y,t)+N_V(x,y,t)$.
%Note that, as defined in the main text, $|\psi(t)\rangle=e^{2\gamma t}|\psi'(t)\rangle$, where $|\psi(t)\rangle$ and $|\psi'(t)\rangle$ are respectively the time-evolved states under $U$, and under the experimentally implemented dynamics $U'$.

We compare the ideal theoretical distribution with
the measured distribution via the similarity,
\begin{equation}
S(t)=\sum_{x,y}\sqrt{P_\text{th}(x,y,t)P_\text{exp}(x,y,t)},
\end{equation}
which quantifies the equality of two probability distributions.
Here $S=0$ stands for completely orthogonal distributions, and $S=1$ for identical distributions. We observe $S\geq 0.914$ in Fig.~\ref{fig:fig2}, $S\geq 0.922$ in Fig.~\ref{fig:fig3}, and $S\geq 0.930$ in Fig.~\ref{fig:fig4}, respectively. Here the theoretical value $P_\text{th}(x,y,t)$ is given by
\begin{align}
P_\text{th}(x,y,t)=\sum_{m=H,V}\frac{|(\langle \bm{r}|\otimes \langle m| )e^{-(\gamma_x+\gamma_y)t}|\psi(t)\rangle|^2}{\sum_{\bm{r},m} |(\langle \bm{r}|\otimes\langle m|)e^{-(\gamma_x+\gamma_y)t}|\psi(t)\rangle|^2},
\end{align}
where $|\psi(t)\rangle$ is the time-evolved walker state under $U$.

In our experiment, photon loss is caused by the loss of photons through an optical element. Our round-trip single-loop efficiency is about $0.66$ even for a unitary quantum walk. This is calculated by multiplying the transmission rates of each optical component used in the round trip, including the transmission rates of the BS ($\sim 0.97$), the collection efficiency from free space to fiber ($\sim 0.78$), the EOM ($\sim 0.96$), and all other optical components ($\sim 0.91$). We therefore estimate the single-loop efficiency as $ 0.78\times0.97\times0.96\times0.91\simeq 0.66$.

{\bf Floquet topological invariant.}
The walker state evolves according to
\begin{align}
|\psi(t)\rangle=U^{t}|\psi(0)\rangle=e^{-iH_{\rm{eff}}t}|\psi(0)\rangle,
\end{align}
where $H_{\rm{eff}}=i\ln U$ is defined as the effective Hamiltonian. While the quantum walk is identified as the periodically-driven Floquet dynamics, the eigenenergies of $H_{\rm{eff}}$ constitute the quasienergy spectrum of the Floquet system. We fix the branch cut of the logarithm such that the quasienergy spectrum lies within the range $[-\pi,\pi)$.

To calculate the Floquet topological invariant, we follow Refs.~\cite{sai,asb,rud}, and define
\begin{equation}
U'=e^{i \tilde{E}}M_{y} S_{y} P' C M_{x} S_{x} C,
\end{equation}
where $e^{i \tilde{E}}$ shifts the quasienergy spectrum by $-\tilde{E}$. The modified phase-shift operator is
\begin{equation}
P'(\beta, \alpha)=\sum_{\bm{r}}\exp \left[i \sigma_{z}(\beta\lfloor x / q\rfloor+2\pi\alpha x)\right]\otimes\ket{\bm{r}}\bra{\bm{r}},
\end{equation}
where $\alpha=p/q$,  $\beta = 2\pi/s$, and $\lfloor x / q\rfloor$ is the greatest integer less than or equal to $x/q$. Here $p$ and $q$ are coprime integers, and $s$  is a sufficiently large integer (in our case, for $\alpha=1/3$, $s=15$ is sufficient).

%To calculate the Floquet topological invariant, we rewrite the Floquet operator as
%\begin{equation}
%U'=e^{i \tilde{E}}M_{y} S_{y} P' C M_{x} S_{x} C,
%\end{equation}
%where $e^{i \tilde{E}}$ can be regarded as shifts the quasienergy spectrum by$-\tilde{E}$,

%bands

We denote the eigenvalues of $U'$ as $e^{-iE_j}$, and the topological invariant for the quasienergy gap (corresponding to $U$) comprising $\tilde{E}$ can be calculated through
\begin{equation}
R=\frac{1}{2 \pi}\left(\sum_{j=1}^{2 s q} E_{j}(1 / s, \alpha, \tilde{E})-\sum_{j=1}^{2 s q} E_{j}(0,\alpha, \tilde{E})\right).
\label{eq:eqR}
\end{equation}
Under an open boundary condition, the value of $R$ indicates the number of anomalous Floquet edge states emerging within the quasienergy gap.

%Without loss of generality, we take $p=1$,$q=3$,$s=15$, $\gamma_{x}=\gamma_{y}=0$ for example, we have $R_{12}=-1, R_{23}=1, R_{45}=-1, R_{56}=1$, where $R_{ij}$ is the \text{RLBL} invariant for different gaps of the quasienergy spectrum.

In Fig.~\ref{fig:fig2}{\bf d}, all quasienergy gaps are closed, hence there are no topological edge states along the boundaries, and the gap topological invariants are ill-defined.
In Fig.~\ref{fig:fig4}, for the $x\leq 0$ ($x>0$) region of the domain-wall configurations,  we have $p=1,q=20$ ($p=-1,q=20$) in Figs.~\ref{fig:fig4}{\bf a, b, c}, and $p=1,q=3$ ($p=-1,q=3$) in Figs.~\ref{fig:fig4}{\bf d, e, f}, respectively. While there are now a host of quasienergy gaps, for any given gap, the topological invariants $R$ of the two regions are always finite and differ by a sign~\cite{supp}. As a consequence, Floquet topological edge modes emerge along the domain-wall boundary.

We find that the Floquet topological invariant $R$ is capable of predicting the anomalous topological edge states under all our experimental parameters, despite the presence of the NHSE. Whether the NHSE can have significant impact on $R$ beyond our experimental parameters (particularly when the photon loss is further increased) is an interesting theoretical question that we leave to future studies.

%In addition, when we reverse the sign of $p$, the \text{RLBL} invariants have the same value but opposite sign, indicating the emergence of edge states with the domain-wall configuration in Fig.~\ref{fig:fig4}.

%\begin{figure}
%\includegraphics[width=0.5\textwidth]{figureS1}
%\caption{}
%\label{fig:figs1}
%\end{figure}

%{\bf Data availability}

%Experimental data, any related experimental background information not mentioned in the text and other findings of this study are available from the corresponding author upon reasonable request.

{\bf Acknowledgments}

We thank Chen Fang for helpful discussions. This work has been supported by the National Natural Science Foundation of China (Grant Nos. 92265209, 12025401 and 11974331). W. Y. acknowledges support from the National Key Research and Development Program of China (Grant Nos. 2016YFA0301700 and 2017YFA0304100).

%{\bf Author contributions}

%Q. L. performed the experiments. W. Y. developed the theoretical aspects and performed the theoretical analysis, and wrote part of the paper. P. X. supervised the project, designed the experiments, analyzed the results and wrote part of the paper.

%{\bf Competing interest declaration}

%The authors declare no competing interests.

%{\bf Additional information}

%Correspondence and requests for materials should be addressed to Wei Yi (wyiz@ustc.edu.cn) and Peng Xue (gnep.eux@gmail.com)

\clearpage

\renewcommand{\thesection}{\Alph{section}}
\renewcommand{\thefigure}{S\arabic{figure}}
\renewcommand{\thetable}{S\Roman{table}}
\setcounter{figure}{0}
\renewcommand{\theequation}{S\arabic{equation}}
\setcounter{equation}{0}

\begin{widetext}

\section{Supplemental Material for ``Manipulating non-reciprocity in a two-dimensional magnetic quantum walk''}

In this Supplemental Material, we provide numerical results characterizing the non-Hermitian skin effect (NHSE) and the Floquet topological invariants.

\subsection{Quasienergy spectra}

The presence of NHSE can be confirmed by examining the quasienergy spectra of the effective Hamiltonian under different boundary conditions, and the spatial distribution of eigenstates under the open boundary condition. As discussed in the main text (Methods section), we define the effective Hamiltonian $H_{\rm{eff}}=i\ln U$, where $U$ is the Floquet operator in Eq.~(1) of the main text, and the branch cut of the logarithm is taken to be the negative real axis.
The quasienergy spectrum of $H_{\rm{eff}}$ thus lies within the range $[-\pi,\pi)$.

In Figs.~\ref{fig:figs1}{\bf a} and {\bf c}, we show the quasienergy spectra under both the periodic (grey) and open (red) boundary conditions. The collapse of the spectra, when the boundary condition is changed from periodic to open, strongly suggests the presence of the NHSE. This is explicitly confirmed in Figs.~\ref{fig:figs1}{\bf b} and {\bf d}, where we show the spatial distribution of the eigenstates under the open boundary condition. They accumulate to one edge or one corner, depending on the loss parameters. Note that $H_{\rm{eff}}$ and $U$ share the same set of eigenstates.

\begin{figure*}[bp]
\centering
\includegraphics[width=0.66\textwidth]{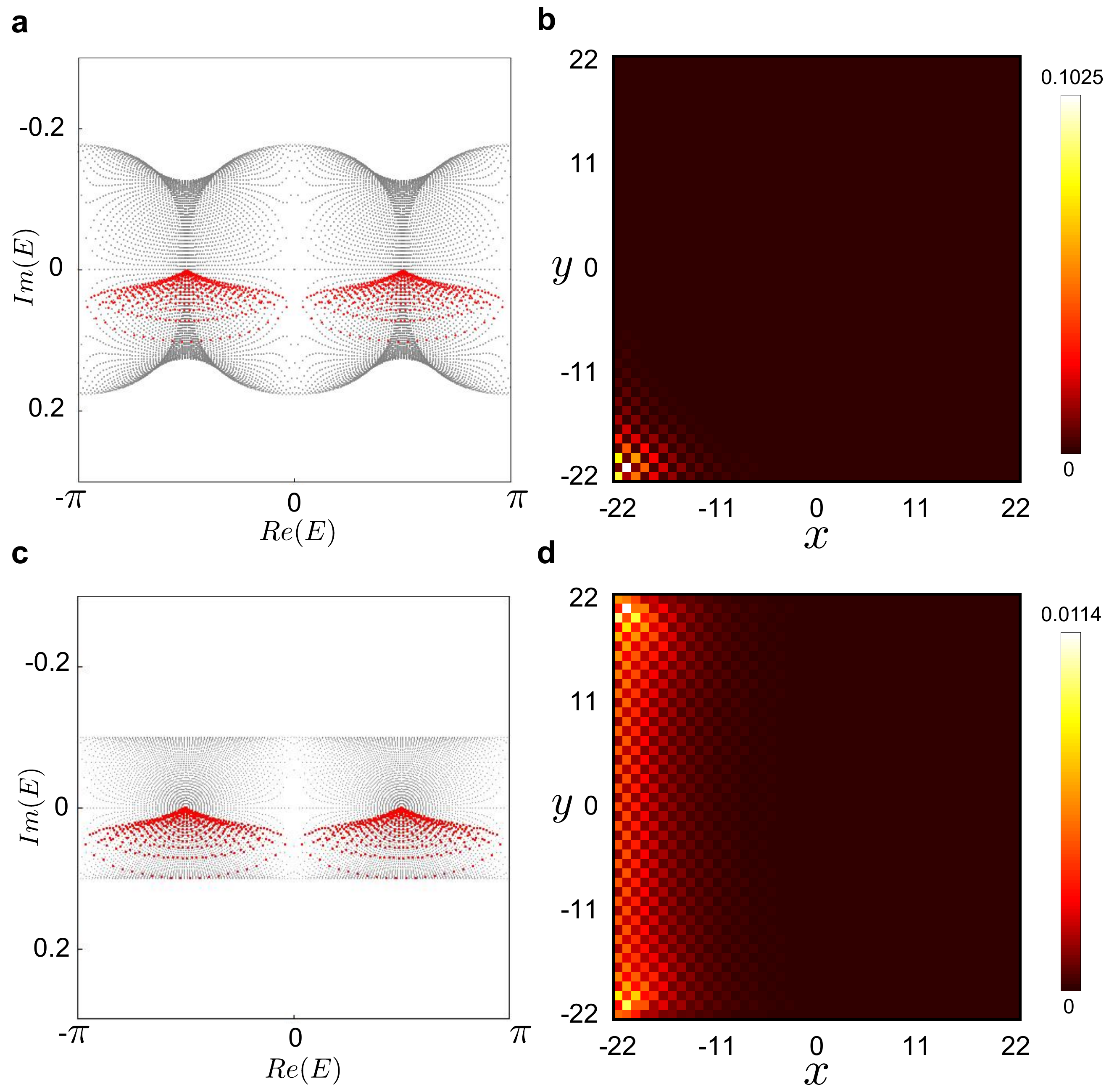}
\caption{{\bf a} Numerically calculated quasienergy spectra of $H_{\rm{eff}}$ on the complex plane,
with $\gamma_{x}=\gamma_{y}=0.125$, under either the periodic boundary condition (grey) or the open boundary condition (red).
{\bf b} Spatial distribution of the eigenstates under the open boundary condition, with the same parameters used in {\bf a}.
{\bf c} Numerically calculated quasienergy spectra of $H_{\rm{eff}}$
with $\gamma_{x}=0.1$, $\gamma_y=0$, under the periodic boundary condition (grey) and the open boundary condition (red), respectively.
{\bf d} Spatial distribution of the eigenstates under the open boundary condition, with the same parameters used in {\bf c}.
For numerical calculations, a lattice size of $45\times45$ is taken as an example, with $\alpha=0$.
}
\label{fig:figs1}
\end{figure*}

\subsection{Floquet topological invariants and edge states}

In Figs.~\ref{fig:figs2}{\bf a} and {\bf b}, we show the real components of the quasienergy spectra under the domain-wall geometry, respectively under the parameters of Fig.~\ref{fig:fig4}{\bf d} and Fig.~\ref{fig:fig4}{\bf f} of the main text. While the Floquet topological edge states are visible within each quasienergy gap, the quasienergy spectra are close to each other, because of the smallness of the loss parameters in Fig.~\ref{fig:figs2}{\bf b}.

The Floquet topological edge states can be characterized by the gap invariant $R$ defined in the Methods section of the main text. In Fig.~\ref{fig:figs2}{\bf c}, we show the calculated gap invariants $R$ for the left (red) and right (black) regions of the domain-wall configuration in Fig.~\ref{fig:figs2}{\bf b}. The topological invariants of the two regions are always finite but differ by their signs. Importantly, within each quasienergy gap, the difference in the gap topological invariants between the two regions is $2$. This corresponds to the number of Floquet topological edge states within each quasienergy gap, as illustrated in Fig.~\ref{fig:figs2}{\bf b}.

\begin{figure*}[tbp]
\centering
\includegraphics[width=\textwidth]{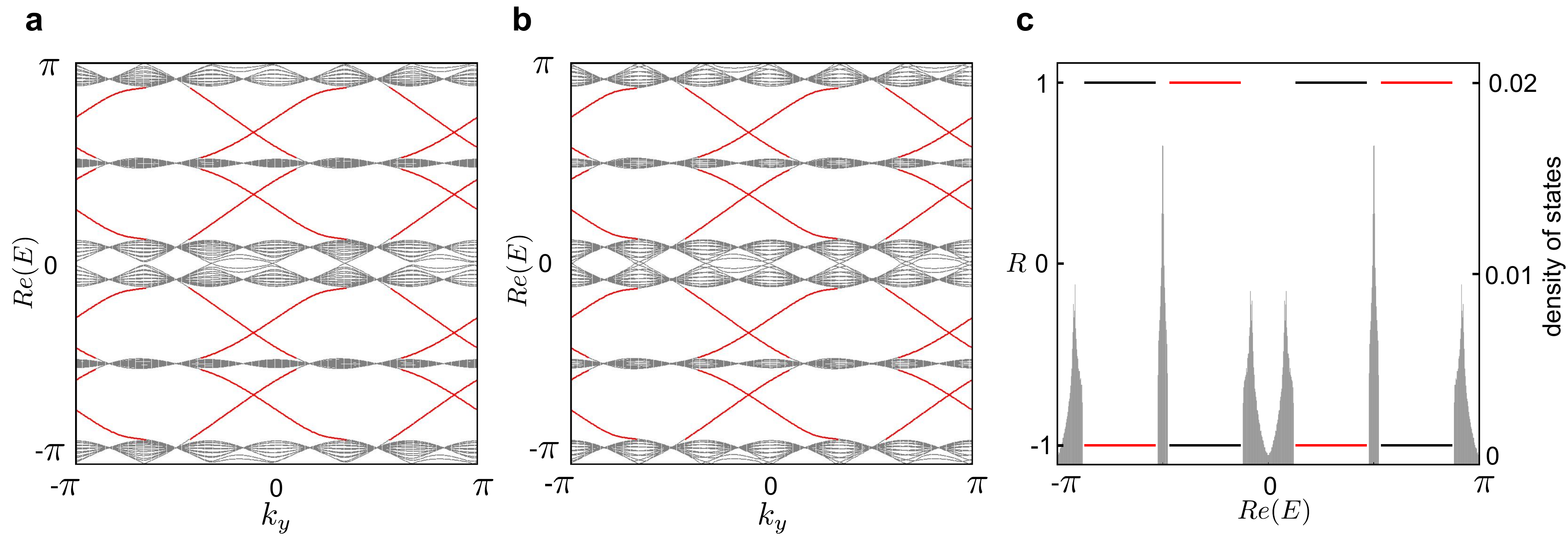}
\caption{{\bf a, b} The real component of the quasienergy spectra under the domain-wall geometry of Figs.~\ref{fig:fig4}{\bf d} and {\bf f} in the main text, with $\gamma_x=\gamma_y=0$ in {\bf a}, and $\gamma_x=\gamma_y=0.1$ in {\bf b}. Floquet topological edge states (red) are seen to emerge within each quasienergy gap.
{\bf c} Floquet topological invariants $R$ as functions of the real component of the quasienergy.
Within each quasienergy gap, the red (black) line indicates $R$ of the left (right) region.
The shaded areas are the density of states normalized to the total number of states, indicating the regions of quasienergy bands.
For numerical calculations, we take a lattice with $45\times45$ sites, and fix $\alpha=1/3$ and $\alpha=-1/3$ for the left and right regions, respectively.
}
\label{fig:figs2}
\end{figure*}

\end{widetext}
\end{document}